
%

\def\leaderdot{\leaders\hbox to 1 em {\hss.\hss}\hfill}

\dimen0= \parindent
\dimen1= \hsize \advance\dimen1 by -\dimen0

\dimen2=\baselineskip
\def\skiplines#1 { \dimen3=\dimen2 \multiply\dimen3 by #1 \vskip \dimen3}
\def\fullline{\hbox to \fullhsize}

\def\numpage{\baselineskip=24pt\fullline{\the\footline}}

\def\mathcedilla{\vtop{\hbox{c}{\kern0pt\nointerlineskip}
	         {\hbox{$\mkern-2mu \mathchar"0018\mkern-2mu$}}}}

\mathchardef\gq="0060
\mathchardef\dq="0027
\mathchardef\ssmath="19
\mathchardef\aemath="1A
\mathchardef\oemath="1B
\mathchardef\omath="1C
\mathchardef\AEmath="1D
\mathchardef\OEmath="1E
\mathchardef\Omath="1F
\mathchardef\imath="10
\mathchardef\fmath="0166
\mathchardef\gmath="0167
\mathchardef\vmath="0176


\def\alignement{\offinterlineskip\halign}
\def\colleft{\strut\kern.3em}
\def\colright{\kern0pt}
\def\filetvide{height2pt}
\def\figureh{\hbox to}

\def\m@th{\mathsurround=0pt}
\newif\ifdtpt
\def\displ@y{\openup1\jot\m@th
    \everycr{\noalign{\ifdtpt\dt@pfalse
    \vskip-\lineskiplimit \vskip\normallineskiplimit
    \else \penalty\interdisplaylinepenalty \fi}}}
\def\eqalignl#1{\,\vcenter{\openup1\jot\m@th
                \ialign{\strut$\displaystyle{##}$\hfil&
                              $\displaystyle{{}##}$\hfil&
                              $\displaystyle{{}##}$\hfil&
                              $\displaystyle{{}##}$\hfil&
                              $\displaystyle{{}##}$\hfil\crcr#1\crcr}}\,}
\def\eqalignnol#1{\displ@y\tabskip\centering \halign to \displaywidth{
                  $\displaystyle{##}$\hfil\tabskip=0pt &
                  $\displaystyle{{}##}$\hfil\tabskip=0pt &
                  $\displaystyle{{}##}$\hfil\tabskip=0pt &
                  $\displaystyle{{}##}$\hfil\tabskip=0pt &
                  $\displaystyle{{}##}$\hfil\tabskip\centering &
                  \llap{$##$}\tabskip=0pt \crcr#1\crcr}}
\def\leqalignnol#1{\displ@y\tabskip\centering \halign to \displaywidth{
                   $\displaystyle{##}$\hfil\tabskip=0pt &
                   $\displaystyle{{}##}$\hfil\tabskip=0pt &
                   $\displaystyle{{}##}$\hfil\tabskip=0pt &
                   $\displaystyle{{}##}$\hfil\tabskip=0pt &
                   $\displaystyle{{}##}$\hfil\tabskip\centering &
                   \kern-\displaywidth\rlap{$##$}\tabskip=\displaywidth
                   \crcr#1\crcr}}
\def\eqalignc#1{\,\vcenter{\openup1\jot\m@th
                \ialign{\strut\hfil$\displaystyle{##}$\hfil&
                              \hfil$\displaystyle{{}##}$\hfil&
                              \hfil$\displaystyle{{}##}$\hfil&
                              \hfil$\displaystyle{{}##}$\hfil&
                              \hfil$\displaystyle{{}##}$\hfil\crcr#1\crcr}}\,}
\def\eqalignnoc#1{\displ@y\tabskip\centering \halign to \displaywidth{
                  \hfil$\displaystyle{##}$\hfil\tabskip=0pt &
                  \hfil$\displaystyle{{}##}$\hfil\tabskip=0pt &
                  \hfil$\displaystyle{{}##}$\hfil\tabskip=0pt &
                  \hfil$\displaystyle{{}##}$\hfil\tabskip=0pt &
                  \hfil$\displaystyle{{}##}$\hfil\tabskip\centering &
                  \llap{$##$}\tabskip=0pt \crcr#1\crcr}}
\def\leqalignnoc#1{\displ@y\tabskip\centering \halign to \displaywidth{
                  \hfil$\displaystyle{##}$\hfil\tabskip=0pt &
                  \hfil$\displaystyle{{}##}$\hfil\tabskip=0pt &
                  \hfil$\displaystyle{{}##}$\hfil\tabskip=0pt &
                  \hfil$\displaystyle{{}##}$\hfil\tabskip=0pt &
                  \hfil$\displaystyle{{}##}$\hfil\tabskip\centering &
                  \kern-\displaywidth\rlap{$##$}\tabskip=\displaywidth
                  \crcr#1\crcr}}


\def\doublelow#1{\,\vtop{\ialign{\hfil$##$\hfil\crcr
                 \mathstrut #1 \crcr}}\,}
\def\charlvmidlw#1#2{\,\vtop{\ialign{##\crcr
      #1\crcr\noalign{\kern1pt\nointerlineskip}
      $\hfil#2\hfil$\crcr}}\,}
\def\charlvlowlw#1#2{\,\vtop{\ialign{##\crcr
      $\hfil#1\hfil$\crcr\noalign{\kern1pt\nointerlineskip}
      #2\crcr}}\,}
\def\charlvmidup#1#2{\,\vbox{\ialign{##\crcr
      $\hfil#1\hfil$\crcr\noalign{\kern1pt\nointerlineskip}
      #2\crcr}}\,}
\def\charlvupup#1#2{\,\vbox{\ialign{##\crcr
      #1\crcr\noalign{\kern1pt\nointerlineskip}
      $\hfil#2\hfil$\crcr}}\,}

\def\vspce{\kern4pt} \def\hspce{\kern4pt}    

\def\emptybox{\vbox{\kern.7ex\hbox{\kern.5em}\kern.7ex}}
 \font\sevmi  = cmmi7              
    \skewchar\sevmi ='177          
 \font\fivmi  = cmmi5              
    \skewchar\fivmi ='177          
\font\tenmib=cmmib10
\newfam\bfmitfam

\textfont\bfmitfam=\tenmib
\scriptfont\bfmitfam=\sevmi
\scriptscriptfont\bfmitfam=\fivmi


\def\twodot{.\kern-0.1em.}

\def\paral{\mathrel{/\kern-.25em/}}
\def\grlo{\mathrel{\hbox{\lower.2ex\hbox{\rlap{$>$}\raise1ex\hbox{$<$}}}}}
\def\logr{\mathrel{\hbox{\lower.2ex\hbox{\rlap{$<$}\raise1ex\hbox{$>$}}}}}
\def\greq{\mathrel{\hbox{\lower1ex\hbox{\rlap{$=$}\raise1.2ex\hbox{$>$}}}}}
\def\loeq{\mathrel{\hbox{\lower1ex\hbox{\rlap{$=$}\raise1.2ex\hbox{$<$}}}}}
\def\grsim{\mathrel{\hbox{\lower1ex\hbox{\rlap{$\sim$}\raise1ex\hbox{$>$}}}}}
\def\losim{\mathrel{\hbox{\lower1ex\hbox{\rlap{$\sim$}\raise1ex\hbox{$<$}}}}}
\font\ninerm=cmr9
\def\uniset{\rlap{\ninerm 1}\kern.15em 1}

\def\emptysq{\mathbin{\vbox{\hrule\hbox{\vrule height1ex \kern.5em
                            \vrule height1ex}\hrule}}}
\def\emptyrect{\mathbin{\vbox{\hrule\hbox{\vrule height1ex \kern1em
                              \vrule height1ex}\hrule}}}
\def\rightonleftarrow{\mathrel{\hbox{\raise.5ex\hbox{$\rightarrow$}\ignorespaces
                                   \lower.5ex\hbox{\llap{$\leftarrow$}}}}}
\def\leftonrightarrow{\mathrel{\hbox{\raise.5ex\hbox{$\leftarrow$}\ignorespaces
                                   \lower.5ex\hbox{\llap{$\rightarrow$}}}}}

\def\bkB{{\rm I\kern-.17em B}}
\def\bkC{{\rm \kern.24em
            \vrule width.05em height1.4ex depth-.05ex
            \kern-.26em C}}
\def\bkD{{\rm I\kern-.17em D}}
\def\bkE{{\rm I\kern-.17em E}}
\def\bkF{{\rm I\kern-.17em F}}
\def\bkG{{\rm \kern.24em
            \vrule width.05em height1.4ex depth-.05ex
            \kern-.26em G}}
\def\bkH{{\rm I\kern-.22em H}}
\def\bkI{{\rm I\kern-.22em I}}
\def\bkJ{{\rm \kern.19em
            \vrule width.02em height1.5ex depth0ex
            \kern-.20em J}}
\def\bkK{{\rm I\kern-.22em K}}
\def\bkL{{\rm I\kern-.17em L}}
\def\bkM{{\rm I\kern-.22em M}}
\def\bkN{{\rm I\kern-.20em N}}
\def\bkO{{\rm \kern.24em
            \vrule width.05em height1.4ex depth-.05ex
            \kern-.26em O}}
\def\bkP{{\rm I\kern-.17em P}}
\def\bkQ{{\rm \kern.24em
            \vrule width.05em height1.4ex depth-.05ex
            \kern-.26em Q}}
\def\bkR{{\rm I\kern-.17em R}}
\def\bkT{{\rm \kern.24em
            \vrule width.02em height1.5ex depth 0ex
            \kern-.27em T}}
\def\bkU{{\rm \kern.30em
            \vrule width.02em height1.47ex depth-.05ex
            \kern-.32em U}}
\def\bkZ{{\rm Z\kern-.32em Z}}



%
\def\midfig#1x#2:#3#4{\midinsert
$$\vbox to #2{\hbox to #1{\special{ps:#3}\hfill}\vfill}$$\par
#4\par\endinsert}
\def\topfig#1x#2:#3#4{\topinsert
$$\vbox to #2{\hbox to #1{\special{ps:#3}\hfill}\vfill}$$\par
#4\par\endinsert}
\def\infig#1x#2:#3{
$$\vbox to #2{\hbox to #1{\special{ps:#3}\hfill}\vfill}$$}
\def\textfig#1x#2:#3{
$\vbox to #2{\hbox to #1{\special{ps:#3}\hfill}\vfill}$}

\input definit.tex
\magnification=1200
\baselineskip=22.0 truept
\centerline{{\bf Eigenmoments for Multifragmentation}}

\bigskip
\centerline{B.G.Giraud and R.Peschanski}
\centerline{\it{Service Physique Th\'eorique, DSM-CE Saclay, F91191
Gif/Yvette, France}}

\bigskip
Abstract: Linear rate equations are used to describe the cascading decay of
an initial heavy cluster into fragments. Using a procedure inspired by
the similar, but continuous case of jet fragmentation in QCD, this
discretized process may be analyzed into
eigenmodes, corresponding to moments of the distribution of multiplicities.
The orders of these moments are usually noninteger numbers. The
resulting analysis can be made time independent and
is applicable to various phenomenological multifragmentation processes,
in which case it leads to new approximate finite-size
scaling relations for the spectrum of fragments.
\bigskip
In this work we consider binary fragmentation processes where any
fragment with mass number $k$ breaks into fragments with mass numbers
$j$ and $k-j,$ $j =1,2...k-1,$ with a probability $w_{j k}$
per unit of time.
It is assumed that $w_{j k}$ is time independent. By definition,
$w_{j k}$
is symmetric if $j$ is replaced by $k-j,$ naturally. Let
$N_j(t)$ be the multiplicity of fragment $j$ at time $t$ in a process initiated
from the decay of a cluster $A,$ namely $N_j(0)=\delta_{jA}.$
The model under study is described by the following set of linear,
first order differential equations,
$$
{dN_j \over dt}= - c_j N_j + \sum_{k=j+1}^A w_{jk} N_k,\ j=1,...A, \eqno (1)
$$
with
$$
c_j=\sum_{\ell=1}^{j-1} {w_{\ell j} \over 2}. \eqno (2)
$$
With components $N_j,\ j=1,...A,$ for a column vector
$|{\cal N}>,$ the system, Eqs.(1),
obviously boils down to
$d|{\cal N}> / dt = {\cal W |N>}$ with a triangular matrix $\cal W.$ For
the sake of clarity we show here $\cal W$ when $A=4,$
$$ \cal W={ \left( \matrix{\displaystyle 0 & \displaystyle w_{12} &
\displaystyle w_{13} & \displaystyle w_{14} \cr\displaystyle 0 &
\displaystyle -w_{12}/2 & \displaystyle w_{23}=w_{13} & \displaystyle w_{24}
\cr\displaystyle 0 & \displaystyle 0 & \displaystyle -(w_{13}+w_{23})/2 &
\displaystyle
w_{34}=w_{14} \cr \displaystyle 0 & \displaystyle 0 & \displaystyle 0 &
\displaystyle
-(w_{14}+w_{24}+w_{34})/2 \cr} \right) }. \eqno (3)
$$
The general solution of Eqs.(1) is obviously a sum of exponentials whose rates
of decay in time are the diagonal matrix elements $-c_j,$ the trivial
eigenvalues of the triangular $\cal W.$ (We notice that any increase of the
dimension $A$ leaves intact the
preexisting eigenvalues and only adds new ones.)

This matrix $\cal W$ has a remarquable property, namely a fixed left
(row-like) eigenstate ${\cal M}_1$, whose components
${\cal M}_{1j}=j,\ j=1,...A,$ do not depend on the $w_{\ell k}$'s.
This comes from the symmetries of $\cal W$ demanded by the conservation of
the total mass $M_1=\sum_{j=1}^A j N_j$, with
$dM_1 / dt=<{\cal M}_1|{\cal W}|{\cal N}>.$
The corresponding eigenvalue
is, naturally, $-c_1=0.$ Moreover, it is clear that the other eigenvalues,
$-c_{\lambda},\ \lambda=2,...A,$ induce a triangular matrix
of ${\rm left}^{\dag}$
eigenstates ${\cal M}_\lambda$, namely
the components ${\cal M}_{\lambda j}$ vanish
when $j<\lambda.$

The higher $\lambda,$
the more the corresponding (bra-like) eigenstate is
only probing heavy fragments in the multiplicity distribution $\cal N$.
This is reminiscent of a hierarchy of moments
$M_q=\sum_{j=1}^A j^q N_j$,
when the exponent $q$ increases. Even though moments
do not make a strictly triangular rearrangement of the information contained
in $\cal N,$ they represent a natural continuation of the first eigenvector
${\cal M}_1,$ and this letter will show that there is a practical connection
between moments and eigenvectors. Indeed, besides triangularity, there is
another argument indicating the interest of moments in the solution of rate
equations like Eqs.(1). In field theory models of jet fragmentation at
high energy, similar evolution equations appear as solutions of perturbative
QCD; they are {\it continuous}, since the quantity which fragments is
the energy-momentum of quarks and gluons, before their transformation into
the observed hadrons. These equations are well known as
Gribov-Lipatov-Altarelli-Parisi
(GLAP) equations$^{[1]}$. As a function of the fraction $x$ of energy-momentum,
the equations for the fragmentation function $N(x)$ can be exactly diagonalized
by
moments $M_q = \int dx x^q N(x).$
In this case, all values of $q\geq 1$ are
admitted, as a consequence of the continuous character of the
equations. Our task, in this letter, is to examine the effect
of discretization on this field-theoretical result. Note that
such a discretization have been introduced long ago$^{[2]}$, but only
for a numerical approximation of the continuous equations.

In the following, we will
consider ``triangularly redefined'' moments
$$
M_{\lambda}=\sum_{j=\lambda}^A j^{q(\lambda)} N_j, \eqno (4)
$$
where the set of exponents $q(\lambda),\ {\lambda}=2,...A,$
may contain nonintegers.
The time derivative of such a redefined moment is
$$
{dM_\lambda \over dt}= \sum^ A_{j=\lambda} j^q \left( \sum^
A_{k=j+1}w_{jk}N_k-c_jN_j \right)= \sum^ A_{k=\lambda} k^qN_kd(\lambda ,k),
\eqno (5)
$$
with
$$
d(\lambda ,k)= \left( \sum^ k_{j=\lambda} \left({j \over k}
\right)^qw_{jk}- {1 \over 2} \sum^ k_{j=1}w_{jk} \right), \eqno (6)
$$
where we have used an interchange of indices $j$ and $k$ in the
 double summation,
and also the
convention that diagonal rates $w_{kk}$ are identically vanishing.
The double summation is sketched on Fig.1, displaying the weight matrix
 $w_{jk}$.

It turns
out that, in practice, there exists special values $q(\lambda )$ for which
the coefficients $d(\lambda,k)$ happen to be, at least approximately,
independent of $k$
when $k$ is large enough. Then, they can be factored out in formula (5)
and they appear as eigenvalues for the corresponding eigenvector $M_\lambda$.
Moreover, simultaneously, since we know the exact eigenvalues (see Eq.(2)), one
gets
$d(\lambda,k) \simeq -c_{\lambda}.$ This will be shown both
analytically in a ``continuous'' limit
(i.e. large matrices) and
numerically for various models with finite sizes $A.$ In fact one is led to
consider first the discrete models which admit the field-theoretical type
of equations in the continuous limit - let us call them the {\it
scale-invariant
 case,}
 since no dependence on the matrix size appears explicitely,
and then the more general situation.

1. {\it The scale-invariant case}

When $A$ is large, $\lambda$ finite, and $k$ large but smaller than $A,$
then the ratio $x=j/k$ can be considered as a continuous label,
$0 \le x \le 1,$
in Eq.(6). Moreover, with positive values of $q$,
an extension of the first summation in Eq.(6), $\sum_{j=\lambda}^k,$ into
a summation $\sum_{j=1}^k$ brings a weak contribution from $0<x<\lambda /k.$
{}From QCD, where $x$ corresponds to the fragmentation of momentum and
a scale-invariant property of the transition weights is valid, one may consider
a large class of models setting
$$
w_{jk}=\varphi(j/k)/k=\varphi(x)/k, \eqno(7)
$$
where $\varphi$ is any suitable function of the scaling
variable $x.$ More precisely, because of the symmetry necessary for $w_{jk},$
a large class of legitimate models correspond to
$w_{jk}=[f(j/k)+f(1-j/k)]/(2k)=[f(x)+f(1-x)]/(2k).$  Hence, for large values
of $k,$ both summations in Eq.(6) amount to the discretization of an integral
$$
d({\lambda})=\int_0^1 dx {[f(x)+f(1-x)](x^q-1/2) \over 2}
=\int_0^1 dx  {[f(x)+f(1-x)][x^q+(1-x)^q-1] \over 4}, \eqno(8)
$$
where $dx$ replaces $1/k$. It will be noticed that $d(\lambda)$ does not
depend any more on $k$. It still
depends on $\lambda$ via the exponent $q(\lambda ),$ naturally. Note also
that the ``splitting'' function $f$ can be general, even with some singularity
at both ends of the integration domain, provided the integral itself
converges.

There may also be a continuous limit for $c_{\lambda}$ if $\lambda$ becomes
large. Indeed, according to Eq.(2),
$ c_{\lambda} \rightarrow \int_0^1 dx \varphi(x)/2, $
if $1/\lambda$ amounts to $dx$
and if this integral converges.
In such a case, the spectrum of $\cal W$ accumulates into a quasi degeneracy.
However, it is important to realize that the convergence of the $c_j$'s
is not required. On the contrary, it is quite possible that the limiting
continuous model does not exist, leading to an infinite hierarchy of
$q(\lambda)$. We shall meet such cases later on. Moreover,
for low values of the label $\lambda,$ this continuous limit
is not in order.

We now notice that $\varphi(x)$ is a semi-positive definite function
since $w_{jk},$ a transition rate, cannot
become a negative number. Hence $d(\lambda)$
is a monotonically decreasing function of $q$. It vanishes for $q=1,$
as expected from the conservation of $M_1.$ According to Eq.(5), the time
evolution of a triangular moment $M_{\lambda}$ becomes very simple if
$d(\lambda )$ can be identified with the eigenvalue, $-c_{\lambda}.$ Hence,
for each $\lambda$, we consider the exponent $q(\lambda )$ which
is the unique solution of the consistency equation
$$
\int_0^1 dx [f(x)+f(1-x)][1-x^{q(\lambda )}-(1-x)^{q(\lambda )}]=4c_{\lambda}.
\eqno (9)
$$
The discrete set of solutions $q_{\lambda}$ of this equation, when the
integer label $\lambda$ runs from $1$ to $A,$
define the ``eigenmoments'' of the theory,
namely those moments whose time evolution is (almost) proportional to just one
exponential $exp(-c_{\lambda}t),$ rather than a mixture of such exponentials.
It will be noticed that, since the sequence of coefficients $c_{\lambda}$
increases monotonically, the sequence of solutions $q(\lambda)$ is
also a monotonically increasing sequence, starting from $q(1)=1$ with $c_1=0.$

Set temporarily $f(x)=1/x^{\beta},$
with $\beta=1.$ This case is reminiscent of the QCD evolution equations for
gluons whose kernel contains the same singularity at small $x$ $^{[1]}$.
Then $w_{jk}=[1/j+1/(k-j)]/2,$ and one finds easily that $c_2=1/2,$
and that the corresponding solution of Eq.(9) is $q(2)=2.$ One also finds
that $c_3=3/4,$ and that the corresponding solution of Eq.(9) is $q(3)=3.$ More
generally, one finds for Eq.(9) the solution $q(\lambda )=\lambda .$
 This definitely
suggests that integer moments form an infinite sequence of
eigenmoments for that choice of $f,$ $f(x)=1/x.$
In agreement with this hint, ones finds easily from Eq.(6) that the sequence
of coefficients $d(2,20)=-0.475,$ $d(2,21)=-0.476,$...$d(2,49)=-0.4898,$
$d(2,50)=-0.4900,$... converges towards $-c_2.$
Just to give another example,
the sequence $d(5,20)=-0.980,$ $d(5,21)=-0.983,$...$d(5,49)=-1.016,$
$d(5,50)=-1.017,$... converges towards $-c_5=-1.042.$ And so on for
all the moments $M_{\lambda},$ which thus generate excellent approximations
to eigenvectors when the exponents $q(\lambda)$ are just integers.

This argument is
independent from the normalization of
$f,$ since Eq.(9) is homogeneous with respect to trivial
multiplications of $f$ by an overall constant.
It may be interesting to note that the diverging
sequence of values for $q(\lambda)$ might be related to the divergence of the
eigenvalue sequence at infinite $\lambda.$

\medskip
A similar result can be observed if $\beta \neq 1,$ but now the solutions of
Eq.(9) do not correspond to integer exponents. For instance, with $\beta=-0.5,$
one finds $c_2=0.177,$ $q(2)\simeq 3.13,$ $c_3=0.232,$ $q(3)\simeq 5.26,$
$c_4=0.259,$ $q(4)\simeq 7.4,$ $c_5=0.275,$ $q(5)\simeq 9.6,$...
The convergence of the coefficients $d$ towards the eigenvalues is still
surprisingly good. For instance $d(2,20)=-0.176,$ and $d(5,20)=-0.275.$

 This
discussion (for the {\it scale invariant} case) is illustrated on Figs.2.
Here, rather than asking whether a moment may behave like an eigenvector,
we consider the reverse question: given an eigenvector $\cal M_{\lambda}$,
does it happen that the components ${\cal M}_{\lambda j}$
induce effective
moments $M_{q(\lambda)}?$ Namely, is there an exponent $q(\lambda)$ compatible
with ${\cal M}_{\lambda j} \propto j^{q(\lambda)},\ j > \lambda ?$ In Figs.2,
 for instance,
the components of the second ($\lambda = 2$) and fourth ($\lambda = 4$)
eigenvectors are displayed as functions of the fragment size. It will be
 stressed that they are almost linear in a Log-Log plot in a large interval
starting from the maximal choosen value $A = 30$. These figures, Figs.2, show
the structure of  moments depending on the parameter $\beta$.

In table I
 we show the comparison between the actual eigenvalues $c_{\lambda},$
 (for $\lambda =2 \ {\rm and}\  4$) with those obtained through  Eq.(9)
after the
determination of the effective values $q(\lambda)$ from Figs.(2). The
agreement is pretty good, except perhaps for $\beta \simeq  2,$ which is at the
 borderline of convergence of the integral in Eq.(9).
It can be thus claimed that various choices of $\varphi$ in Eq.(7) make it
possible to find eigenmoments
via the continuous limit and solutions of Eq.(9).
It will be noticed that this continuous limit is closely linked to the
denominator $k$ in Eq.(7), since this denominator induces the needed measure
$dx,$ independently of the overall scale given by $A.$

2. {\it Scale-dependent cases}

In a more general situation, $w_{jk}$ is not compatible with Eq.(7). This will
in general lead to an introduction of the overall scale $A$
in the problem, and corresponds to cases where the scale invariance
of the weights is not preserved. Let us illustrate this by the following
instance:
 $$
w_{jk}=[(j/k)^{-\beta}+(1-j/k)^{-\beta}]/(2k^{\alpha}), \eqno (10)
$$
where $\alpha$ may be different from $1$.

To be specific, but as an example of more general value, we display
on Fig.3 the Log-Log plot ${\cal M}_{\lambda j}$ versus $j$,
for $\lambda = 2, A=50, \beta = 0$ and various values of $\alpha$.
 From this figure,
one realizes on this simple example that the diagonalization by
eigenmoments is obtained for $\alpha \le 1,$ while for $\alpha > 1$
there is a clear distorsion of eigenvectors with respect to moments. It is
interesting to interpret this phenomenon analytically by inspecting the
modification which occurs with the choice of
 Eq.(10) for the weights. Instead of Eqs.(5-6), one finds the following,
$$
{dM_\lambda \over dt}= \sum^ A_{k=\lambda} k^{q-\alpha +1} N_k d(\lambda ,k),
\quad d(\lambda ,k)= \left( \sum^ k_{j=\lambda} \left({j \over k}
\right)^q\bar w_{jk}- {1 \over 2} \sum^ k_{j=1}\bar w_{jk} \right), \eqno (11)
$$
where $\bar w$ corresponds to {\it rescaled} weights with
$\alpha =1.$ Following the previous discussion, based on the existence
of a continuous limit $d(\lambda,k) \simeq d(\lambda),$ one  is led to
the approximate consistency equation:
$$
-d(\lambda) \simeq A^{\alpha -1} c(\lambda), \eqno (12)
$$
where the renormalisation factor $A^{\alpha -1}$ takes care of the
initial values of the moments. The occurrence of the $A-$dependent
factor is the signal of the lack of scale invariance of the fragmentation
dynamics when $\alpha \neq 1.$ Note that Eq.(12) uses the
assumption that the eigenvalues do not change substantially between
$q$ and $q-1+\alpha.$ It can only be an approximation.

In Table II, we display the different values obtained
for the $q(2)$ for various values of $\alpha,$ obtained
by the fitted slopes at the origin for the curves obtained in
Fig.3. We compare them to those obtained from Eq.(12)
when the input are the actual exact eigenvalues $c_2.$ The agreement
is here also quite satisfactory, except in the region when
$\alpha > 1$. One might associate this phenomenon to the
well-known$^{[3]}$ fact that a shattering transition takes place at
finite time, the conservation of mass being broken in the continuum limit.
A special study of this case
is in order for the future.

\bigskip
The class of models which
can be analyzed by eigenmoments is thus larger than the class described by
Eq.(7). It must be noted, however, that the ``eigenorders'' $q(\lambda)$
are not universal, but clearly model-dependent.

As an application of the properties of eigenmoments, let us consider
the problem of 3-dimensional bond percolation in a finite-size square lattice.
This model seems to give a successful description of nuclear
multifragmentation, when a heavy ion receives enough excitation energy
to form a highly unstable state and decays into several fragments$^{[4]}$.
The statistics of fragment numbers and sizes seem to follow predictions
of a percolation model in which each lattice site is populated by a
nucleon, and the percolation parameter $p,$ namely the survival probability
for
bonds, varies between 0 and 1. There is no obvious time
scale in this model, hence a comparison of its predictions with
those of linear rate equations models requires the use of eigenmoments
in order to obtain an intrinsic time scale from the evolution of such
eigenmoments.

For this purpose, we remark that, if they are identified
as eigenmoments, the $M_q$'s
are linked by linear relations in Log-Log plots, and their
explicit time dependence disappears. We are thus
led to display in the same way the moments obtained
from the percolation model, choosing for instance $M_2$
for reference, see Fig.4. Different moments are displayed
(with $q= 1,1.5,2,3,4,5$) and show the interesting
feature of a quasi-linear dependence for the values $q=3,4,5,$ given the
fact that for $q=1$ (mass conservation) and $q=2$ (reference scale) the linear
dependence is fixed. It is clear from this figure that the quasi-linear form
is obtained between $p=1$ and $p=p_c,$ where $p_c$ is the critical value above
which, in the continuous limit, an {\it infinite} percolation cluster is
formed. Indeed, the figure shows
the dominant contribution of the cluster of largest
mass to the averaged moments.
This largest cluster  is, for finite size problems,
the representative of the infinite
cluster when $p \ge p_c.$

Notice that the moments implied by the rate equations
are the {\it full} moments, including the largest fragment, while in usual
analyses of percolation models$^{[4]}$, scaling properties are
investigated with
moments modified by the subtraction of the largest cluster.
Moreover, in such traditional analyses of percolation, the reference time
scale is generally given by the moment $M_0$ or a similar variable
related to the multiplicity of fragments. The comparison and
compatibility of our approach with such analyses is an open
problem of some interest$^{[6]}$.

{\it In conclusion,} from this first study on the percolation
model, we obtain a hint that linear rate equations
could provide a time dependent
description of multifragmentation. But it is difficult
at this stage to obtain informations on the set of eigenorders
$q(\lambda)$ which
could be associated with percolation.
The existence of scaling relations between moments can be
proven in the vicinity of $p_c$ for percolation through finite-size
scaling$^{[4]}$. Our result is compatible with this and, furthermore,
involves the whole region
$1>p>p_c.$ In the representation provided by rate equations, we have obtained
scaling relations valid for
short time scales, while the previous general results, see
for instance Ref.$^{[7]}$, involve long time scales only. An open
problem is to connect both analyses for a general
system of equations.

{\it Acknowledgments } R.P. thanks Xavier Campi for his patient explanations
on nuclear multifragmentation and for providing the authors with the suitable
percolation
program. Thanks are due to G\' erard Auger, Brahim Elattari, Pierre Grang\' e,
Hubert Krivine, Eric Plagnol and Jean Richert for fruitful discussions.

$^{\dag}Footnote$ The right-hand-side, ket-like eigenstates define also
a triangular matrix, apparently unrelated to the matrix of bra eigenstates,
except for
trivial biorthogonality relations. Up to now we have been unable to find
a practical use of these ket-like eigenvectors.

\medskip
\centerline{{\bf References}}
\medskip

\item{[1]} G. Altarelli and G. Parisi, {\it Nucl. Phys.}{\bf 126} (1977) 297.
 V.N. Gribov and L.N. Lipatov, {\it Sov. Journ. Nucl. Phys.}{\bf 15} (1972) 438
and 675.
For a review and references, {\it Basics of perturbative QCD} Y.L. Dokshitzer,
V.A. Khoze, A.H. Mueller and S.I. Troyan (J. Tran Than Van ed. Editions
Fronti\` eres, France, 1991.)

\item{[2]} P. Cvitanovic, P. Hoyer and K. Zalewski, {\it Nucl. Phys. B}
{\bf 176} (1980) 429.

\item{[3]} E.D. Mc Grady and Robert M. Ziff
{\it Phys. Rev. Lett.} {\bf 58} (1987) 892.

\item{[4]} X. Campi, {\it Phys. Lett. B}
{\bf 208} (1988) 351, and contributions to the
the Proceedings of Varenna 1990 and 1992
Summer Courses of the International School of Physics {\it Enrico Fermi}.

\item{[5]} For a general review on percolation:
D. Stauffer, {\it Introduction to Percolation Theory}
(Taylor and Francis, London and Philadelphia, Penn. 1985.)

\item{[6]} On a phenomenological ground in relation to nuclear
multifragmentation, the problem has been
raised by: J. Richert and P. Wagner, {\it Nucl. Phys. A}{\bf 517} (1990) 299.

\item{[7]} Z. Cheng and S. Redner,
{\it J. Phys. A: Math. Gen.} {\bf 23} (1990) 1233.

\vfill
\eject

\centerline { {\bf Figure captions} }
\medskip

\noindent
{\bf Fig.1: Discrete rate equations: Double Summation}

\noindent
The double summation range for truncated moments $M_{\lambda}$
is represented by the hatched triangle. White dots: diagonal weigths $w_{jj}$
are zero. Black dots: non-diagonal weigths $w_{jk}$. The interchange of
indices $j$ and $k$ in the description of
the hatched triangle leads to Eqs.(5-6).

\noindent
{\bf Fig.2: Eigenvectors: scale-invariant case $(\alpha = 1)$}

\noindent
The components of the rate equation eigenvectors $M_{\lambda j}$ are displayed
as functions of $j$ in a Log-Log plot. The chosen weights correspond to
Eq.(10) with $\alpha = 1,$ and different values of $\beta.$ The curves
correspond to a smooth interpolation (dark line), resp. extrapolation (dashed
line), of the exact eigencomponents for a system of size $A = 30.$
{\it Fig 2-a: $\lambda = 2$ ; Fig 2-b: $\lambda = 4.$}
The curves are used for the determination of the effective values $q(2), q(4),$
see Table I.

\noindent
{\bf Fig.3: Eigenvectors: scale-dependent cases $(\beta = 0)$}

\noindent
Same as Figure 2-a but for weights following Eq.(10) with $\beta = 0$
and different values of $\alpha.$ The curves give the eigenorders listed
in Table II.

\noindent
{\bf Fig.4 Percolation analyzed with the $``M_2$ time scale''}

\noindent
Relative strengths of moments
$M_q, \  q=0,1,1.5,2,3,4,5,$ as functions of $M_2$
in a Log-Log plot. Data taken from 3-d bond percolation on a $6*6*6$ lattice.
The corresponding values of the bond survival probability $p$ are shown on
the horizontal axis. Its critical value is $p_c = .25.$
Full lines: moments. Dashed lines: contributions of the largest cluster.
Dashed-dotted line: the reference moment $M_2.$
Notice that a linear behaviour is approximately obtained for $0 \le p \losim
p_c$ and $q = 3,4,5.$

\medskip
\centerline {\bf{Table Captions}}

{\bf Table I:} For $\alpha=1, \lambda =2,4$ and
different values of $\beta,$
the effective exponents $q(\lambda)$ from Figs.2-a,b
and the corresponding values
for $-d(\lambda ),$ see Eq.(9). The latter
are compared with the exact eigenvalues $c_{\lambda}.$

{\bf Table II:} For $\beta = 0, \lambda = 2$ and different values of $\alpha,$
comparison of the measured effective eigenorders (obtained
from Fig.3) with those predicted from Eq.(9).

\vfill
\eject

$$ \vbox{\alignement
{& \vrule#& \colleft# \colright
& \vrule#& \colleft# \colright
& \vrule#& \colleft# \colright
& \vrule#& \colleft# \colright
& \vrule#& \colleft# \colright
& \vrule#& \colleft# \colright
& \vrule#& \colleft# \colright
&\vrule#\cr
\noalign{\hrule}
\filetvide& \omit&& \omit&& \omit&& \omit&& \omit&& \omit&& \omit&\cr
&$\displaystyle \hfill \beta \hfill$&&$\displaystyle \hfill q(2)
\hfill$&&$\displaystyle \hfill -d(2) \hfill$&&$\displaystyle \hfill c_2
\hfill$&&$\displaystyle \hfill q(4) \hfill$&&$\displaystyle \hfill -d(4)
\hfill$&&$\displaystyle \hfill c_4 \hfill$&\cr
\filetvide& \omit&& \omit&& \omit&& \omit&& \omit&& \omit&& \omit&\cr
\noalign{\hrule}
\filetvide& \omit&& \omit&& \omit&& \omit&& \omit&& \omit&& \omit&\cr
&$\displaystyle \hfill -2 \hfill$&&$\displaystyle \hfill 2.4
\hfill$&&$\displaystyle \hfill .061 \hfill$&&$\displaystyle \hfill .062
\hfill$&&$\displaystyle \hfill 6.1 \hfill$&&$\displaystyle \hfill .11
\hfill$&&$\displaystyle \hfill .11 \hfill$&\cr
\filetvide& \omit&& \omit&& \omit&& \omit&& \omit&& \omit&& \omit&\cr
&$\displaystyle \hfill -1.5 \hfill$&&$\displaystyle \hfill 2.7
\hfill$&&$\displaystyle \hfill .09 \hfill$&&$\displaystyle \hfill .09
\hfill$&&$\displaystyle \hfill 6.5 \hfill$&&$\displaystyle \hfill .14
\hfill$&&$\displaystyle \hfill .14 \hfill$&\cr
\filetvide& \omit&& \omit&& \omit&& \omit&& \omit&& \omit&& \omit&\cr
&$\displaystyle \hfill -.5 \hfill$&&$\displaystyle \hfill 3.1
\hfill$&&$\displaystyle \hfill .18 \hfill$&&$\displaystyle \hfill .18
\hfill$&&$\displaystyle \hfill 7.4 \hfill$&&$\displaystyle \hfill .26
\hfill$&&$\displaystyle \hfill .26 \hfill$&\cr
\filetvide& \omit&& \omit&& \omit&& \omit&& \omit&& \omit&& \omit&\cr
&$\displaystyle \hfill 0 \hfill$&&$\displaystyle \hfill 3
\hfill$&&$\displaystyle \hfill .25 \hfill$&&$\displaystyle \hfill .25
\hfill$&&$\displaystyle \hfill 7.1 \hfill$&&$\displaystyle \hfill .38
\hfill$&&$\displaystyle \hfill .38 \hfill$&\cr
\filetvide& \omit&& \omit&& \omit&& \omit&& \omit&& \omit&& \omit&\cr
&$\displaystyle \hfill .5 \hfill$&&$\displaystyle \hfill 2.6
\hfill$&&$\displaystyle \hfill .36 \hfill$&&$\displaystyle \hfill .35
\hfill$&&$\displaystyle \hfill 5.8 \hfill$&&$\displaystyle \hfill .57
\hfill$&&$\displaystyle \hfill .57 \hfill$&\cr
\filetvide& \omit&& \omit&& \omit&& \omit&& \omit&& \omit&& \omit&\cr
&$\displaystyle \hfill 1 \hfill$&&$\displaystyle \hfill 2
\hfill$&&$\displaystyle \hfill .51 \hfill$&&$\displaystyle \hfill .50
\hfill$&&$\displaystyle \hfill 4.2 \hfill$&&$\displaystyle \hfill .94
\hfill$&&$\displaystyle \hfill .92 \hfill$&\cr
\filetvide& \omit&& \omit&& \omit&& \omit&& \omit&& \omit&& \omit&\cr
&$\displaystyle \hfill 1.5 \hfill$&&$\displaystyle \hfill 1.5
\hfill$&&$\displaystyle \hfill .79 \hfill$&&$\displaystyle \hfill .71
\hfill$&&$\displaystyle \hfill 2.8 \hfill$&&$\displaystyle \hfill 1.8
\hfill$&&$\displaystyle \hfill 1.6 \hfill$&\cr
\filetvide& \omit&& \omit&& \omit&& \omit&& \omit&& \omit&& \omit&\cr
&$\displaystyle \hfill 2 \hfill$&&$\displaystyle \hfill 1.1
\hfill$&&$\displaystyle \hfill .85 \hfill$&&$\displaystyle \hfill 1
\hfill$&&$\displaystyle \hfill 1.7 \hfill$&&$\displaystyle \hfill 4.6
\hfill$&&$\displaystyle \hfill 2.7 \hfill$&\cr
\filetvide& \omit&& \omit&& \omit&& \omit&& \omit&& \omit&& \omit&\cr
\noalign{\hrule}}}
 $$
\par
\centerline{Table I} \par

\bigskip
$$ \vbox{\alignement
{& \vrule#& \colleft# \colright
& \vrule#& \colleft# \colright
& \vrule#& \colleft# \colright
&\vrule#\cr
\noalign{\hrule}
\filetvide& \omit&& \omit&& \omit&\cr
&$\displaystyle \hfill \alpha \hfill$&&$\displaystyle \hfill \doublelow{ q(2)
\cr measured \cr} \hfill$&&$\displaystyle \hfill \doublelow{ q(2) \cr
predicted \cr} \hfill$&\cr
\filetvide& \omit&& \omit&& \omit&\cr
\noalign{\hrule}
\filetvide& \omit&& \omit&& \omit&\cr
&$\displaystyle \hfill 0 \hfill$&&$\displaystyle \hfill 1.03
\hfill$&&$\displaystyle \hfill 1.04 \hfill$&\cr
\filetvide& \omit&& \omit&& \omit&\cr
&$\displaystyle \hfill .5 \hfill$&&$\displaystyle \hfill 1.17
\hfill$&&$\displaystyle \hfill 1.22 \hfill$&\cr
\filetvide& \omit&& \omit&& \omit&\cr
&$\displaystyle \hfill .6 \hfill$&&$\displaystyle \hfill 1.26
\hfill$&&$\displaystyle \hfill 1.33 \hfill$&\cr
\filetvide& \omit&& \omit&& \omit&\cr
&$\displaystyle \hfill .8 \hfill$&&$\displaystyle \hfill 1.64
\hfill$&&$\displaystyle \hfill 1.7 \hfill$&\cr
\filetvide& \omit&& \omit&& \omit&\cr
&$\displaystyle \hfill 1 \hfill$&&$\displaystyle \hfill 3.0
\hfill$&&$\displaystyle \hfill 3 \hfill$&\cr
\filetvide& \omit&& \omit&& \omit&\cr
&$\displaystyle \hfill 1.1 \hfill$&&$\displaystyle \hfill 5.7
\hfill$&&$\displaystyle \hfill 5.2 \hfill$&\cr
\filetvide& \omit&& \omit&& \omit&\cr
&$\displaystyle \hfill 1.2 \hfill$&&$\displaystyle \hfill 20
\hfill$&&$\displaystyle \hfill 14 \hfill$&\cr
\filetvide& \omit&& \omit&& \omit&\cr
\noalign{\hrule}}}
 $$
\par
\centerline{Table II} \par

\bye